**Melting of the vortex lattice through intermediate hexatic fluid in *a*-MoGe thin film**

Indranil Roy[a], Surajit Dutta[a], Aditya N. Roy Choudhury[a], Somak Basistha[a], Ilaria Maccari[b], Soumyajit Mandal[a], John Jesudasan[a], Vivas Bagwe[a], Claudio Castellani[b], Lara Benfatto[b], and Pratap Raychaudhuri[a][1]

[a]Tata Institute of Fundamental Research, Homi Bhabha Rd, Colaba, Mumbai 400005, India.

[b] ISC-CNR and Department of Physics, Sapienza University of Rome, P. le A. Moro 5, 00185 Rome, Italy.

The hexatic fluid refers to a phase in between a solid and a liquid which has short range positional order but quasi-long range orientational order. In the celebrated theory of Berezinskii, Kosterlitz and Thouless and subsequently refined by Halperin, Nelson and Young, it was predicted that a 2-dimensional hexagonal solid can melt in two steps: first, through a transformation from a solid to a hexatic fluid which retains quasi long range orientational order and then from a hexatic fluid to an isotropic liquid. In this paper, using a combination of real space imaging and transport measurements we show that the 2-dimensional vortex lattice in *a*-MoGe thin film follows this sequence of melting as the magnetic field is increased. Identifying the signatures of various transitions on the bulk transport properties of the superconductor, we construct a vortex phase diagram for a two dimensional superconductor.

---

[1] pratap@tifr.res.in

Ever since in their seminal work[1], Berezinski, Kosterlitz and Thouless (BKT) predicted the possibility of a phase transition without breaking continuous symmetry in 2-dimensional (2D) systems, a lot of effort has been devoted to explore its ramifications in different systems. 2D crystalline solids present an interesting situation. Melting of 3 dimensional crystalline solids is understood through the "Lindemann criterion", where the solid melts through a first order phase transition when the root mean square lattice vibration amplitude exceeds a certain fraction of the lattice constant[2]. In contrast, for a 2D solid, the BKT theory extended subsequently by Halperin, Nelson and Young (HNY), predicted that melting could also proceed through an alternate route[3,4,5], via two continuous phase transitions mediated via topological defects. At the first transition, thermally excited free dislocations proliferate in the lattice creating an intermediate state between a crystalline solid and a liquid. At the second transition, dislocations dissociate into isolated disclinations producing an isotropic fluid. The intermediate state (called a hexatic fluid when the solid has hexagonal symmetry) has zero shear modulus and short-range positional order like in a liquid, but retains the quasi long-range orientational order of the parent solid. Over the years there have been several attempts to test the BKTHNY theory in diverse 2D systems such as electrons over a liquid He surface[6], inert-gas monolayers adsorbed on graphite, vortices in superconducting thin films[7,8,9,10] and colloidal crystals[11,12,13,14]. Indeed, according to the various experimental conditions one can either prove the occurrence of the melting transition at the expected value, or the existence of an orientational order when the translational one is lost, but the simultaneous observation of the two features has so far been available only in the case of colloidal crystals[12,13,14].

In a clean conventional superconductor, the vortices arrange themselves into a hexagonal lattice, known as the Abrikosov vortex lattice (VL)[15,16]. In general, the vortices can meander along its length inside the superconductor, and therefore the VL is not strictly a 2D system. However, in thin films, the thickness of the sample can be orders of magnitudes smaller than the characteristic bending length of the vortices[7]. In this limit, the vortices behave like point objects, and from the standpoint of vortices the VL behaves effectively like a 2D hexagonal solid. The progress in low temperature scanning tunneling spectroscopy (STS), which allows the imaging of the VL over wide range of magnetic field and moderately large areas[17], has triggered efforts to directly observe the hexatic vortex fluid state. However, this simple scenario is complicated by the presence of additional ingredients. First, in crystalline superconducting films the VL can get strongly coupled to the symmetry of the crystalline lattice[18,19,20] thereby influencing its orientational order. Secondly, the presence of crystalline defects and impurities creates a random pinning landscape which can trap vortices at specific locations. While the first complicacy can be avoided in thin films of amorphous superconductors, some degree of random pinning is practically unavoidable.

From a theoretical standpoint, random pinning can easily destroy translational order whereas its effect on orientational order is weaker. At low fields, it is now accepted that in the presence of weak pinning the VL is in a solid-like phase called Bragg glass[21,22,23] which has long range orientational order and quasi-long range positional order. However, at larger pinning density the vortex state can become unstable to dislocations even in the absence of thermal excitations[24,25] producing a hexatic glass. This disorder induced hexatic glass differs from a hexatic fluid from the fact that it has a non-zero shear modulus, and the dislocations are frozen in space[26] unlike thermally generated dislocations in a hexatic fluid which statistically appear at random locations. Thus in general, a 2-D Bragg glass could transform into a hexatic fluid, a hexatic glass or even undergo a first order transition into a vortex liquid. While transport[8] or magnetic shielding measurements[7] can establish the melting of a Bragg glass fairly accurately, establishing the hexatic nature of vortex state above the melting transition is not straightforward. On the other hand, real space imaging such as STS can identify a hexatic state from the quasi long-range orientational order and the nature of topological defects[9,10,27,28,29]; but since dynamics of the vortices in the fluid phase can be extremely slow, it is not easy to distinguish between a hexatic glass and a hexatic fluid.

In this letter, we adopt a strategy which combines magnetotransport and STS imaging to investigate the melting of the VL in a well characterized amorphous MoGe (*a*-MoGe) thin film. We determine the transition from a vortex solid to a vortex fluid from detailed magnetotransport measurements and use STS imaging to identify the nature of the fluid from topological defects and to study the temporal dynamics of the vortices. The central result of this paper is that as the magnetic field is increased the vortex state goes successively from a vortex solid to a hexatic fluid and then to an isotropic liquid following the sequence expected from BKTHNY theory. Using the characteristic features associated with these transitions we construct a phase diagram for the 2D vortex state.

The sample used in this study consists of *a*-MoGe thin films with thickness, $t \sim 20$ nm and $T_c \sim 7.05 \pm 0.05$ K (lower *inset* Fig. 1(a)), grown through pulsed laser deposition. Details of sample growth, characterization and experimental methods is given in the supplementary material[30]. Due to different requirements of shape and size, two samples with the same $T_c$, and thickness variation <10%, were used for transport and STS measurements. For STS, post-deposition, the film was transferred *in-situ* in an ultra-high vacuum suitcase and transferred in the Scanning Tunneling Microscope without exposure to air.

We first investigate the field at which the vortex solid undergoes transition to a vortex fluid. In the mixed state above a critical current, $I_c$, the Lorentz force on the vortices exceeds the pinning force and a flux-flow regime is established. Here, the current-voltage (*I-V*) characteristics follows a linear relation[31], $V = R_{ff}(I-I_c)$, where $R_{ff}$ is the flux flow resistance. However, at any finite temperature, even for $I \ll I_c$, a small but finite voltage appears due to thermally activated flux flow (TAFF) over the pinning barrier, $U$, giving a TAFF resistance, $R_{TAFF} = V/I = R_{ff} \exp(-U/kT)$, (where $k$ is the Boltzman constant). The difference between a vortex solid and a vortex fluid comes from the response in this TAFF region at low currents. For a vortex solid, $U$ depends on current[32,33,22] as $U(I) = U_0(I_c/I)^\alpha$, such that $R_{TAFF}$ (and $V$) exponentially goes to zero for $I \to 0$. In contrast, in a vortex fluid[34] $U$, and hence $R_{TAFF}$, is independent of current at low currents. In Fig. 1 (a) we show $I_c$ in the magnetic field range 0.9 – 5 kOe obtained by fitting the linear flux flow region of the *I-V* curves at 2 K ( Fig. 1(a) *upper inset* ). Fig. 1(b) shows the magnetic field variation of TAFF resistivity, $\rho_{TAFF}$, obtained from the slope of the linear region of the *I-V* curves for $I <$ 100$\mu A \ll I_c$(*inset* Fig. 1(b)), after taking the geometric factor into consideration. We observe that $\rho_{TAFF}$ remains zero up to 1.9 kOe and then increases gradually. Above 1.9 kOe, the near perfect linearity of the *I-V* curves at low currents is characteristic of an ideal vortex fluid. To further confirm that this field indeed corresponds to the vortex solid to vortex fluid transition we investigate the functional form of the *I-V* curves for $I < I_c$ (Fig. 1(c)). Below the transition (1.5 and 1.8 kOe) the *I-V* curves can be fitted very well with the form expected for a vortex solid (eqn. (1)) with $\alpha=1$. In contrast, above the transition (2.2 and 2.5 kOe), the *I-V* curves significantly deviate from the exponential dependence. Instead, at low currents a linear slope appears below $I \approx 200$ µA as expected for a vortex fluid. The temperature dependence of $\rho_{TAFF}$ above and below the melting field is consistent with this scenario[30].

To identify the nature of the vortex fluid we now use STS imaging. To obtain VL images, spatially resolved tunneling conductance ( $G(V) = dI/dV$ ) was measured at different fields using a Pt-Ir tip. The vortex core in a superconductor behaves like a normal-metal where both the gap and the coherence peak in the local density of states are suppressed. Consequently, when the bias voltage ($V$) is kept close to the superconducting coherence peak, each vortex manifests as a local minima[28,29,19] in $G(V)$. Fig. 2(a)-(f) show representative large area VL images at 2 K along with their 2-D Fourier transforms (FT). We obtain the precise position of the vortices from the local minima in the conductance map and identify the topological defects by Delaunay triangulating the VL. At 1 kOe, we observe a hexagonal VL without any topological defects. Above 3 kOe we observe free dislocations in the VL as expected for a vortex fluid. However, up to 70 kOe the FT shows six spots, showing the existence of six-fold orientational order. Above 70 kOe the FT transforms into a ring corresponding to an isotropic vortex liquid. Since topological defects are free to move in a fluid, we expect the defects to appear at different locations when the VL is imaged at different times. To explore this, in Fig. 3, we show three successive images of the VL captured over the same area, where the time to acquire each image is 1.5 hrs. In each image dislocations appear at a different location and sometimes disappear from the field of view. Thus our data are consistent with a hexatic vortex fluid below ~ 70 kOe, and an isotropic vortex liquid at higher fields. Hexatic VL are also observed[30] between 10 - 70 kOe at 450 mK.

An important property of the hexatic fluid is that due to its orientational order, the motion of vortices should happen preferentially along the principal direction of the VL. To explore this we follow the motion of the vortices on a finer scale, by capturing a series of 12 successive images over the same area at 15 minutes intervals (Fig. 4 (a)-(i)). Since no drive is applied here, the motion of the vortices is caused by stress relaxation of the VL. The stress in the VL could in principle be of two kinds: a residual global stress when the VL has not yet reached its true equilibrium or local stress in the lattice caused by the appearance of short lived dislocations. Though in our sample the global stress is unlikely to be an issue due to the very weak pinning nature of the film, as an added precaution we apply several magnetic field pulses of 0.3 kOe before the data is taken. At 1 kOe, in the vortex solid phase, each vortex only undergoes small wandering motion about its mean position (see, Supplementary Material[30]). In the hexatic fluid (3 kOe and 55 kOe), we observe that although the motion of individual vortices is irregular and follows a jagged trajectory, over a longer time-scale all vortices within the field of view preferentially move along one of the three principal directions of the VL. This regularity of motion is disturbed only in the vicinity of locations where topological defects appear, where the movement becomes more random. On the other hand, at 85 kOe the motion becomes completely random as expected for an isotropic fluid.

To quantify the orientational order of the vortex state, we now compute the six-fold orientational order parameter[35], defined as, $\Psi_6 = \frac{1}{N}\langle\sum_{k,l} e^{[6i(\phi_k-\phi_l)]}\rangle$; here $\phi_k$ is the angle between a fixed direction in the plane of the VL and the $k$ - th bond and the sum runs over all the bonds in the VL. $\Psi_6 = 1$ for a perfect hexagonal lattice. Since we are using finite area images, to improve the statistics we average over the series of 12 images. In Fig. 5(a), we plot the magnetic field variation of $\Psi_6$ at 2K. $\Psi_6$ remains finite up to 55 kOe and abruptly drops to a small value above 70 kOe signaling the transition from a hexatic fluid to a vortex liquid[36,30]. On the same plot we show $\rho_{TAFF}$ extracted in the same way as before, for the entire magnetic field range. We observe that above 70 kOe $\rho_{TAFF}$ increases rapidly reflecting the increased mobility of the vortices as the system enters the isotropic vortex liquid state. At low temperatures, we also observe a shallow minimum in $\rho_{TAFF}$ reminiscent of the "peak effect" observed in thicker samples[37]; the origin of this effect is currently unclear and needs to be investigated further.

We can now use the $\rho_{TAFF}$-$H$ variation at different temperatures (Fig. 5(b)) to construct the phase diagram in the $H$-$T$ parameter space (Fig. 5(c)). The upper critical field, $H_{c2}$, is determined from the same graph as the field where the resistivity reaches the normal state value. It is interesting to note that the both solid-hexatic fluid and hexatic fluid-vortex liquid phase boundaries keep increasing in field down to the lowest temperature as expected for the thermal melting transition[8], instead of flattening out at low temperatures as often observed when the order-disorder transition is driven by disorder[22,38]. Further investigation is needed to determine to what extent this phase diagram is generic, for example, in thin crystalline superconductors (such as monolayer[39] NbSe$_2$) or in layered High-$T_c$ cuprates, where the vortex state can be effectively 2-D over a large part of the $H$-$T$ parameter space[40]. Finally, we would also like to note that though we have used magnetic field, or alternatively, the density of vortices, as the tuning parameter in our experiments, one would also expect to observe the two-step melting as a function of temperature. However, it might be more difficult to observe the transition as a function of temperature in imaging experiments, since the contrast in STS images becomes poor at elevated temperatures.

In summary, we have shown a clear demonstration of the BKTHNY type two-step melting of the 2D vortex lattice in a very weakly disordered $a$-MoGe thin film. We believe that the simplicity of the system combined with the ability to investigate the static and dynamic response of the VL using a variety of probes such as high frequency conductivity and precision magnetometry, will pave the way to a more detailed understanding of defect driven phase transitions.

*Acknowledgements:* PR would like to thank Nandini Trivedi and LB would like to thank Thierry Giamarchi and Dragana Popovic for valuable discussion. The work was supported by Department of Atomic Energy, Govt. of India, Department of Science and Technology, Govt of India (Grant No: EMR/2015/000083), and India-Italy joint project grant (No. INT/Italy/P-21/2016 (SP) and MAECI SUPERTOP-PGRO4879).

[1] J. M. Kosterlitz and D. J. Thouless, *Early Work on Defect Driven Phase Transitions*, in 40 years of Berezinskii-Kosterlitz Thouless Theory, edited by J. V. Jose (World Scientific, Singapore, 2013).

[2] F. A. Lindemann, *The Calculation of Molecular Vibration Frequencies*, Phys. Z. **11**, 609 (1910).

[3] B. I. Halperin and D. R. Nelson, *Theory of Two-Dimensional Melting*, Phys. Rev. Lett. **41**, 121 (1978); Phys. Rev. Lett. **41**, 519 (1978)

[4] A. P. Young, *Melting and the vector Coulomb gas in two dimensions*, Phys. Rev. B **19**, 1855 (1979).

[5] V. N. Ryzhov, E. E. Tareyeva, Yu D. Fomin and E. N. Tsiok, *Berezinskii–Kosterlitz–Thouless transition and two-dimensional melting*, Physics-Uspekhi **60,** 857 (2017).

[6] W. F. Brinkman, Daniel S. Fisher, D. E. Moncton, *Melting of Two-Dimensional Solids*, Science **217**, 693 (1982).

[7] A. Yazdani, C. M. Howald, W. R. White, M. R. Beasley and A. Kapitulnik, *Competetition between pinning and melting in the two-dimensional vortex lattice*, Phys. Rev. B **50**, R16117 (1994).

[8] P. Berghuis, A. L. F. van der Slot, and P. H. Kes, *Dislocation-mediated vortex-lattice melting in thin films of a-$Nb_3Ge$*, Phys. Rev. Lett. **65**, 2583 (1990).

[9] I. Guillamón, H. Suderow, A. Fernández-Pacheco, J. Sesé, R. Córdoba, J. M. De Teresa, M. R. Ibarra and S. Vieira, *Direct observation of melting in a two-dimensional superconducting vortex lattice,* Nat. Phys. **5**, 651 (2009).

[10] I. Guillamón, R. Córdoba, J. Sesé, J. M. De Teresa, M. R. Ibarra, S. Vieira, and H. Suderow, *Enhancement of long-range correlations in a 2D vortex lattice by an incommensurate 1D disorder potential,* Nat. Phys. **10**, 851 (2014).

[11] R. E. Kusner, J. A. Mann, J. Kerins, and A. J. Dahm, *Two-Stage Melting of a Two-Dimensional Collodial Lattice with Dipole Interactions*, Phys. Rev. Lett. **73**, 3113 (1994).

[12] K. Zahn, R. Lenke, and G. Maret, *Two-Stage Melting of Paramagnetic Colloidal Crystals in Two Dimensions*, Phys. Rev. Lett. **82**, 2721 (1999).

[13] P. Keim, G. Maret, and H. H. von Grünberg, *Frank's constant in the hexatic phase*, Phys. Rev. E **75**, 031402 (2007).

[14] S. Deutschlander, T. Horn, H. Lowen, G. Maret, and P. Keim, *Two-Dimensional Melting under Quenched Disorder*, Phys. Rev. Lett. **111**, 098301 (2013).

[15] A. A. Abrikosov, *The magnetic properties of superconducting alloys*, J. Phys. Chem. Solids. **2**, 199 (1957).

[16] L. P. Lévy, *Vortices in Type II Superconductors.* in Magnetism and Superconductivity: Texts and Monographs in Physics. Springer, Berlin, Heidelberg (2000).

[17] H Suderow, I Guillamón, J G Rodrigo and S Vieira, *Imaging superconducting vortex cores and lattices with a scanning tunneling microscope*, Supercond. Sci. Technol. **27**, 063001 (2014).

[18] J. Toner, *Orientational order in disordered superconductors*, Phys. Rev. Lett. **66**, 2523 (1991).

[19] S. C. Ganguli, H. Singh, R. Ganguly, V. Bagwe, A. Thamizhavel and P. Raychaudhuri, *Orientational coupling between the vortex lattice and the crystalline lattice in a weakly pinned $Co_{0.0075}NbSe_2$ single crystal*, J. Phys.: Condens. Matter **28**, 165701 (2016).

[20] V. G. Kogan, M. Bullock, B. Harmon, P. Miranovic, Lj. Dobrosavljevic-Grujic, P. L. Gammel, and D. J. Bishop, *Vortex lattice transitions in borocarbides,* Phys. Rev. B **55**, R8693 (1997).

[21] T. Giamarchi and P. Le Doussal, *Elastic theory of flux lattices in the presence of weak disorder,* Phys. Rev. B **52**, 1242 (1995).

[22] T. Giamarchi and P. Le Doussal, *Phase diagrams of flux lattices with disorder,* Phys. Rev. B **55**, 6577 (1997).

[23] T. Giamarchi T. *Disordered Elastic Media*, In: Meyers R. (eds) Encyclopedia of Complexity and Systems Science. Springer, New York, NY (2009). ( DOI: https://doi.org/10.1007/978-0-387-30440-3_127 )

[24] E. M. Chudnovsky, *Structure of a solid film on an imperfect surface,* Phys. Rev. B **33**, 245 (1986); *Hexatic vortex glass in disordered superconductors*, *ibid.* **40**, 11355(R) (1989); *Orientational and positional order in flux lattices of type-II superconductors*, *ibid*. **43**, 7831 (1991).

[25] Min-Chul Cha and H. A. Fertig, *Disorder-Induced Phase Transitions in Two-Dimensional Crystals*, Phys. Rev. Lett. 74, 4867 (1995).

[26] J. P. Rodriguez, *Macroscopic phase coherence of defective vortex lattices in two dimensions*, Phys. Rev. B **72**, 214503 (2005).

[27] M. Zehetmayer, *How the vortex lattice of a superconductor becomes disordered: a study by scanning tunneling spectroscopy*, Sci. Rep. **5**, 9244 (2015).

[28] S. C. Ganguli, H. Singh, G. Saraswat, R. Ganguly, V. Bagwe, P. Shirage, A. Thamizhavel and P. Raychaudhuri, *Disordering of the vortex lattice through successive destruction of positional and orientational order in a weakly pinned $Co_{0.0075}NbSe_2$ single crystal*, Sci. Rep. **5**, 10613 (2015).

[29] S. C. Ganguli, H. Singh, I. Roy, V. Bagwe, D. Bala, A. Thamizhavel and P. Raychaudhuri, *Disorder-induced two-step melting of vortex matter in Co-intercalated NbSe2 single crystals*, Phys. Rev. B **93**, 144503 (2016).

[30] Sample growth and experimental details is given in Section 1 of the Supplementary Material; the basic superconducting characterisation of the MoGe film from superconducting energy gap and penetration depth is given in Section 2; Vortex lattice images at 450 mK are given in Section 3. The spatial variation of the orientational order parameter $G_6(r)$ is discussed in Section 4. The temperature dependence of $\rho_{TAFF}$ is discussed in Section 5. The real space visualisation of vortex creep is shown in Section 6.

[31] T. P. Orlando and K. A. Delin, *Foundations of Applied Superconductivity, Addison-Wesley (1991)*.

[32] M. V. Feigel'man, V. B. Geshkenbein, A. I. Larkin, and V. M. Vinokur, *Theory of collective flux creep*, Phys. Rev. Lett. **63**, 2303 (1989).

[33] D. S. Fisher, M. P. A. Fisher, and D. A. Huse, *Thermal fluctuations, quenched disorder, phase transitions, and transport in type-II superconductors*, Phys. Rev. B **43**, 130 (1991).

[34] V. M. Vinokur, M. V. Feigel'man, V. B. Geshkenbein, and A. I. Larkin, *Resistivity of high-$T_c$ superconductors in a vortex-liquid state,* Phys. Rev. Lett. **65**, 259 (1990).

[35] S. A. Hattel and J. M. Wheatley, *Flux-lattice melting and depinning in the weakly frustrated two-dimensional XY model*, Phys. Rev. B **51**, 11951 (1995).

[36] The small but finite value of $\Psi_6$ observed at 70 kOe results from the large short range orientation correlation length close to hexatic fluid-vortex liquid boundary. For further details, see ref. 30, Section 4.

[37] R. Wordenweber, P. H. Kes and C. C. Tsuei, *Peak and history effects in two-dimensional collective flux pinning*, Phys. Rev. B 33, 3172 (1986).

[38] B. Khaykovich, E. Zeldov, D. Majer, T. W. Li, P. H. Kes, and M. Konczykowski, *Vortex-Lattice Phase Transitions in $Bi_2Sr_2CaCu_2O_8$ Crystals with Different Oxygen Stoichiometry,* Phys. Rev. Lett. **76,** 2555 (1996).

[39] H. Wang et al., *High-quality monolayer superconductor $NbSe_2$ grown by chemical vapour deposition*, Nat. Comm. **8**, 394 (2017).

[40] Z. Shi, P. G. Baity, T. Sasagawa, Dragana Popovic, *Unveiling the phase diagram of a striped cuprate at high magnetic fields: Hidden order of Cooper pairs,* arXiv:1801.06903v1 ( *unpublished* ).

**Figure Captions**

**Figure 1|** (a) $I_c$ vs. $H$ between 0.9 – 5 kOe at 2K. The *upper inset* shows the *I-V* curves at 2 K in different magnetic fields; some curves have been omitted for clarity. The *lower inset* shows the temperature variation of resistivity, ρ, in zero field. (b) $\rho_{TAFF}$ as function of magnetic field calculated from the low current region of *I-V* curves; in the same plot we show the characteristic current, $I_0$, below which $V < 0.1$ μV. The *inset* shows the expanded view of the *I-V* curves below 100 μA. (c) *I-V* curves for $I < I_c$ for 4 fields spanning the vortex solid to vortex fluid transition. The red lines show the fit to the TAFF equation with $U = U_0(I_c/I)$ ($U_0$ is an adjustable parameter). The blue lines are linear fit to the *I-V* curve below 200 μA. The insets of the lower two panels show the expanded view of the linear fit.

**Figure 2|** (a)-(f) Representative vortex images at 2 K for 1 kOe, 3 kOe, 10 kOe, 55 kOe, 70 kOe and 85 kOe respectively. The vortices, (denoted by the black dots) appear as minima in the conductance map recorded at a fixed d.c. bias, $V_b = 1.52$ mV. The VL is Delaunay triangulated to find out the topological defects, which are denoted as red, green, magenta and yellow dots, corresponding to 5, 7, 4 and 8-fold coordination. Above each vortex image is the 2-D Fourier transform (FT) of the image.

**Figure 3|** (a)-(d) Three consecutive vortex images captured one after the other at the same location at 2 K for 3 kOe, 25 kOe, 55 kOe respectively. The color coding of the vortices, as well as, the topological defects are similar to Fig. 2. We observe that dislocations appear at different positions in each image.

**Figure 4|** (a), (b) and (c) First image of 12 consecutive vortex images at 2 K for 3 kOe, 55 kOe and 85 kOe respectively. The color codes for vortices and defects are same as Fig. 4. (d), (e) and (f) are arrow-maps for each fields, where each arrow gives the displacement for every vortex through individual steps of 12 consecutive vortex images. The red boxes in the arrow-maps are zoomed in (g), (h) and (i).

**Figure 5|** (a) Variation of the orientational order parameter $\Psi_6$ as a function of magnetic field at 2 K. $\rho_{TAFF}$ measured at the same temperature is shown in the same plot. A representative error bar on $\Psi_6$ is shown on 40 kOe data point. The vertical dashed line demarcates the hexatic fluid from the vortex liquid. (b) $\rho_{TAFF}$ vs. $H$ at different temperatures; the horizontal dashed line denotes normal state resistivity. (c) Phase diagram showing the vortex solid, hexatic fluid and vortex liquid phase in the *H-T* parameter space. The phase boundary between vortex solid and hexatic fluid is multiplied by 10 on the magnetic field axis for clarity.

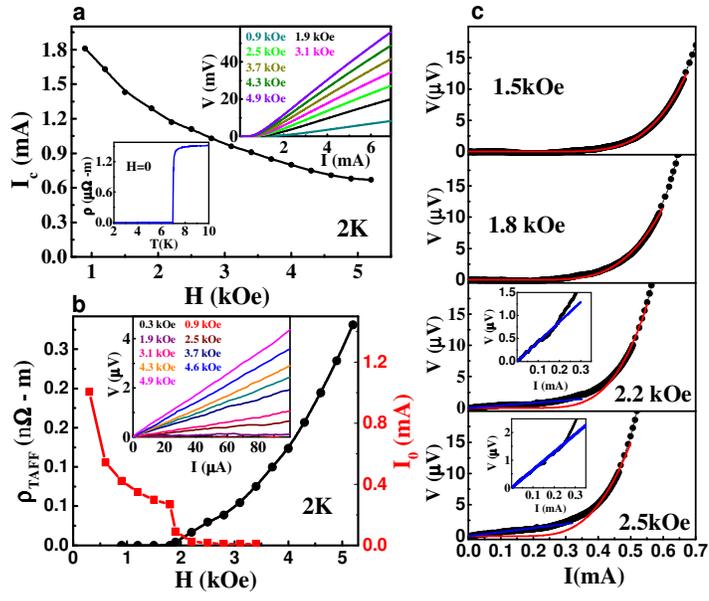

**Figure 1**

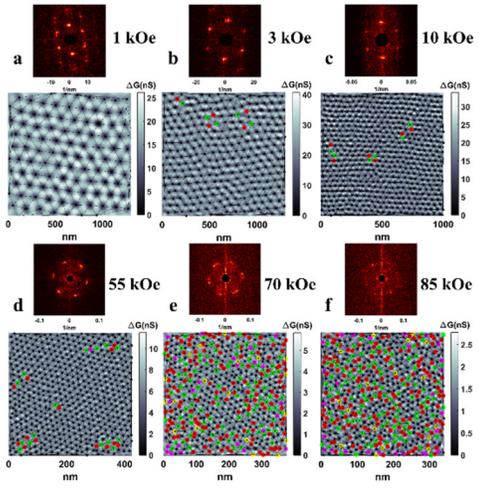

**Figure 2**

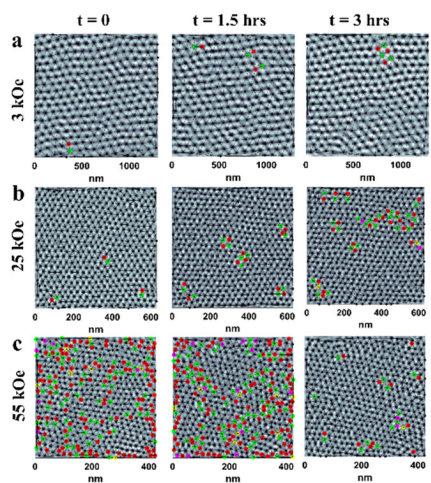

**Figure 3**

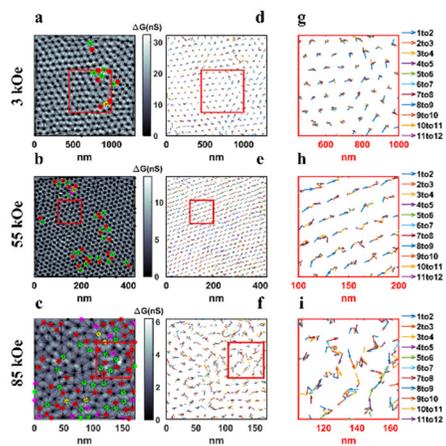

**Figure 4**

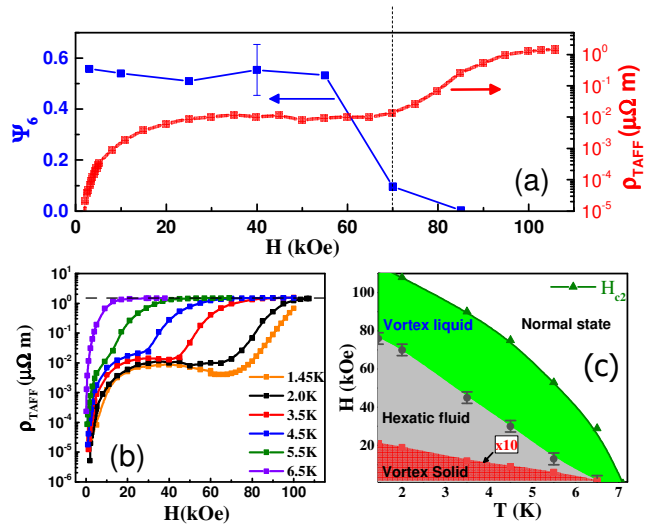

**Figure 5**

**Supplementary material: Melting of the vortex lattice through intermediate hexatic fluid in *a*-MoGe thin film**

Indranil Roy[a], Surajit Dutta[a], Aditya N. Roy Choudhury[a], Somak Basistha[a], Ilaria Maccari[b], Soumyajit Mandal[a], John Jesudasan[a], Vivas Bagwe[a], Claudio Castellani[b], Lara Benfatto[b], and Pratap Raychaudhuri[a,2]

[a] Tata Institute of Fundamental Research, Homi Bhabha Rd, Colaba, Mumbai 400005, India.

[b] ISC-CNR and Department of Physics, Sapienza University of Rome, P. le A. Moro 5, 00185 Rome, Italy.

**Section 1. Sample growth and experimental details**

*Sample growth:* The sample used in this study consists of *a*-MoGe thin film with thickness, $t \sim 20$ nm grown on surface oxidized Si substrate through pulsed laser deposition. The amorphous film was synthesized by ablating a $Mo_{70}Ge_{30}$ bulk target using a 248 nm excimer laser keeping the substrate at room temperature. The deposition is carried out in vacuum of $1\times10^{-8}$ Torr. Since for a metallic thin film a comparatively high energy density is needed to maintain the stoichiometry close to the stoichiometry of the target (compared to oxides) the laser was focused in a tight spot with repetition rate of 10 Hz on the target, giving an effective energy density $\sim 240$ mJ/mm$^2$ per pulse. The growth rate was $\sim 1$ nm/100 pulse. Scanning electron micrographs revealed a flat and featureless surface with very low density of particulates. Energy dispersive X-ray analysis on the film gave a stoichiometry of $Mo_{71\pm1.5}Ge_{29\pm1.5}$. The sample used for magnetotransport measurements was capped with a 2 nm thick Si layer to prevent surface oxidation. . For STS, post deposition, the film was transferred *in-situ* in an ultra-high vacuum suitcase (base pressure $10^{-10}$ Torr) and transferred in the STM without exposure to air.

---

[2] pratap@tifr.res.in

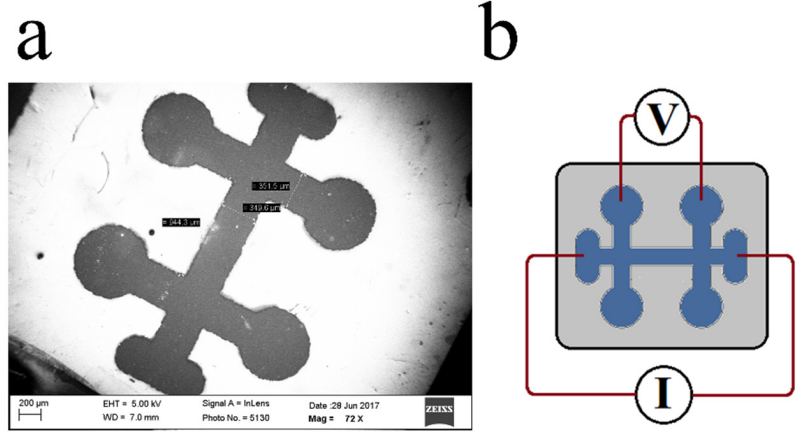

*Figure 1S: (a) Patterned sample for 4-probe magnetotransport measurements, where width of the bridge is 0.3 mm and length is 1.3 mm. (b) Schematic diagram of the 4-probe magnetotransport measurement using the patterned MoGe sample.*

*Magnetotransport measurements:* Magnetotransport measurements were carried out using standard 4-probe technique in a $^4$He cryostat using a current source and a nanovoltmeter. To improve the sensitivity for both current-voltage (*I-V*) characteristics and resistance (*R*) measurements, the film was patterned as shown in Fig. 1S (a) and the current was passed across a 1.3 mm long and 0.3 mm wide bridge and the voltage was measured across the bridge (Fig. 1S (b)).

*Low frequency penetration depth:* We measured the complex penetration depth $\lambda_\omega$ using a low-frequency two-coil mutual inductance technique, where an 8 mm diameter *a*-MoGe film of the same thickness and $T_c$ as the ones used for transport and STS imaging, was sandwiched between a primary coil and a secondary coil (*inset* of Fig. 2S(b)). The inductive and dissipative components of the mutual inductance ($M = M' - iM''$) between the two were measured by passing a small 31 kHz a.c. current in the primary and measuring the out-of-phase and in-phase component of the voltage in the secondary using a lock-in-amplifier. To determine $\lambda_\omega$ from $M$ we use the following method. $\lambda_\omega^{-2}$ can be decomposed into the magnetic penetration depth, $\lambda$ and the skin depth δ, as, $\lambda_\omega^{-2} = \lambda^{-2} + i\delta^{-2}$. We create a lookup table by calculating $M$ for a range of $\lambda^{-2}$ and $\delta^{-2}$, by numerically solving the Maxwell and London equations using finite element analysis. The experimental values of $\lambda$ and $\delta$ are obtained by comparing the calculated values of $M$ with the measured ones. Details of this technique are described in ref. 41, 42, 43.

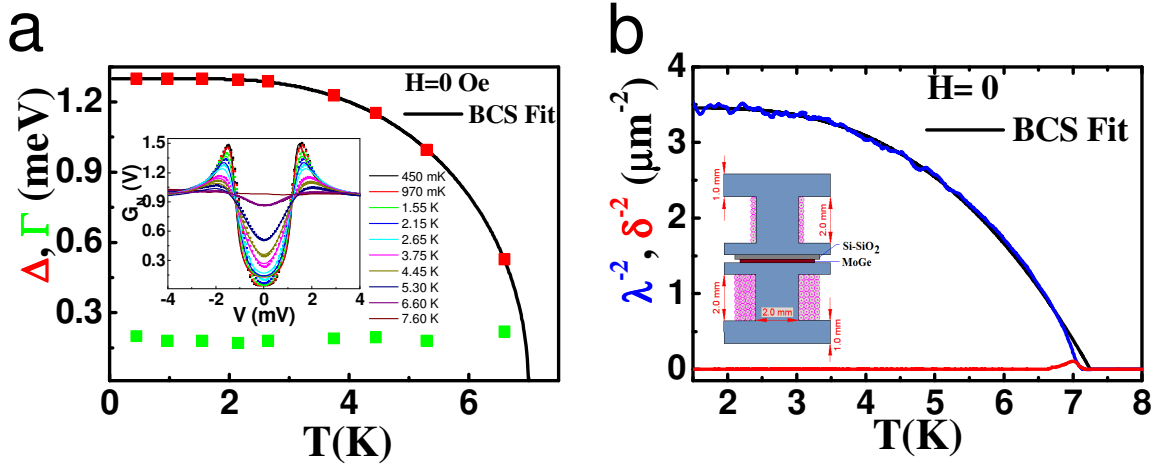

***Figure 2S:*** *(a) Temperature variation of the superconducting energy gap, Δ (red squares), and phenomenological broadening parameter Γ (green squares) obtained by fitting the $G_N(V)=G(V)/G(4\ mV)$ vs V spectra shown in the inset; the fits to the tunneling spectra are shown in the inset. The black line is the expected variation of Δ(T) obtained from BCS theory. (b) Temperature variation of $\lambda_L^{-2}$ and $\delta^{-2}$ in zero field, where $\lambda_L$ is the low frequency magnetic penetration depth. The variation fits well with dirty limit BCS formula, except very close to $T_c$, giving a low temperature value of $\lambda_L \sim 534$ nm. (inset) Schematic diagram of the two-coil probe used for ac screening response measurements. The top quadrupolar coil is the primary and the bottom dipolar one is the secondary and the sample is sandwiched between the two.*

*Vortex lattice imaging using STS:* We imaged the VL using a home-built low temperature scanning tunneling microscope[44] (STM) operating down to 350 mK and fitted with a 90 kOe superconducting solenoid. To obtain VL images, spatially resolved tunneling conductance (*dI/dV*) was measured at different field using a Pt-Ir tip. The vortex core in a superconductor behaves like a normal-metal where both the gap and the coherence peak in the local density of states are suppressed. Consequently, when the bias voltage ( *V* ) is kept close to the superconducting coherence peak, the position of each vortex manifests as a local minima in *dI/dV*. We obtain the precise position of the vortices by determining the local minima in the conductance map and identify the topological defects by Delaunay triangulating the VL. In addition, we study the temporal dynamics of the vortices by taking several successive images at fixed time intervals and tracking the motion of each vortex.

**Section 2: Basic characterization of superconducting properties of the a-MoGe film**

The superconducting transition temperature, $T_c$, and the temperature variation of the upper critical field, $H_{c2}$ are presented in the main body of the paper. Here we present the temperature variation of the superconducting energy gap and the London penetration depth. The superconducting energy gap, $\Delta$, is determined from low temperature STM measurements in zero field (Fig. 2S(a)). $\Delta$ is determined by fitting the zero field tunneling spectra (*inset* Fig. 2S(a)) with the tunneling equation[45], $G(V) = (1/R_N) \int_{-\infty}^{\infty} N_s(E) \left(-\frac{\partial f(E-eV)}{\partial E}\right) dE$, where *f(E)* is the Fermi-Dirac distribution function, *e* is the electron charge, $R_N$ is the tunneling resistance at high bias ($V \gg \Delta/e$) and $N_s(E)$ is the single particle density of states (DOS) in the superconductor. The spectra fit well using the Bardeen-Cooper-Schrieffer (BCS) expression, $N_s(E) = Re\left(\frac{|E|+i\Gamma}{\sqrt{(|E|+i\Gamma)^2 - \Delta^2}}\right)$, where the additional broadening parameter $\Gamma$ phenomenologically takes into account broadening of the DOS from non-thermal origin[46]. We obtain $\Delta(0) \approx 1.3$ meV, and $\Delta(T)$ follows the expected BCS variation for an s-wave superconductor. The London penetration depth, $\lambda_L$ ($\approx \lambda_\omega$), is determined from two-coil measurements in zero applied d.c. magnetic field. For a conventional superconductor in the dirty limit, the temperature variation of $\lambda_L^{-2}$ follows the BCS formula[45], $\frac{\lambda_L^{-2}(T)}{\lambda_L^{-2}(0)} = \frac{\Delta(T)}{\Delta(0)} tanh\left[\frac{\Delta(T)}{k_B T}\right]$, where $k_B$ is the Boltzman constant. We obtain a good fit to the data (Fig. 2S (b)) using $\Delta(0) \approx 1.36$ meV (which is close to the tunneling gap) and a low temperature value of $\lambda_L(0) \approx 534$ nm. As expected in zero field $\delta^2$ is zero except for a small peak close to $T_c$.

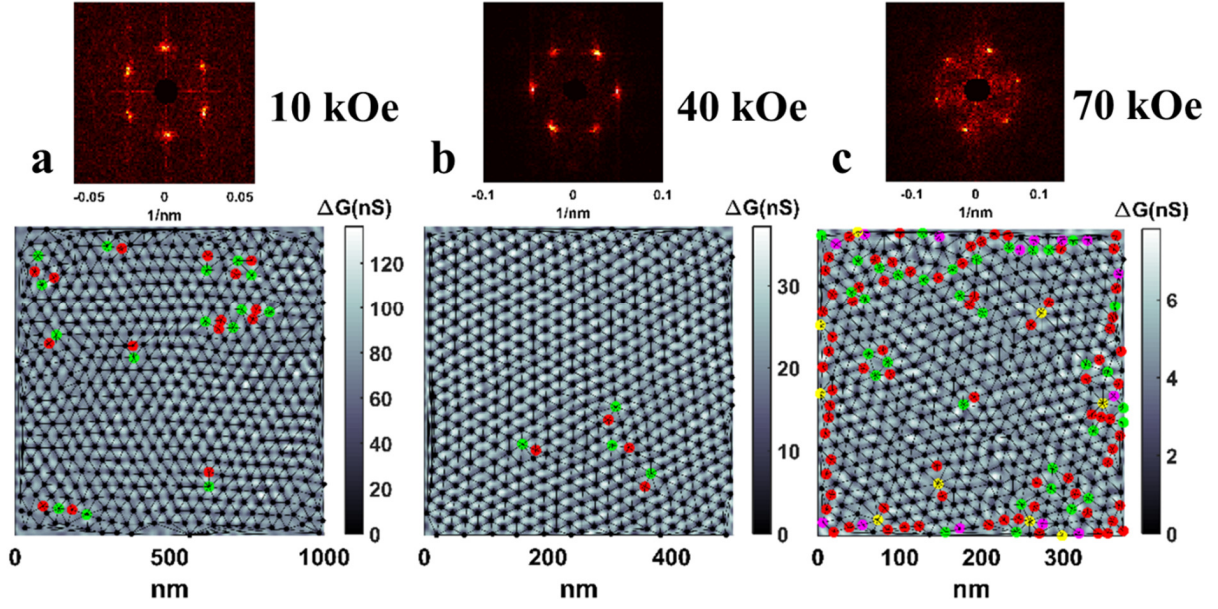

*Figure 3S:* *(a)-(c) Representative vortex images at 450 mK for 10 kOe, 40 kOe and 70 kOe. The vortices, (denoted by the black dots) appear as minima in the conductance maps recorded at a fixed d.c. bias, $V_b$ = 1.52 mV. The VL is Delaunay triangulated to find out the topological defects, which are denoted as red, green, magenta and yellow dots, corresponding to 5, 7, 4 and 8-fold coordination. Above each vortex image is the 2-D Fourier transform (FT) of the image.*

**Section 3: Real space imaging of the hexatic fluid at 450 mK**

While most STS imaging was performed at 2K, we also verified the existence of the hexatic state at 450 mK. In Fig. 3S(a)-(c) we show the Delaunay triangulated vortex lattice images along with the corresponding 2-D Fourier transforms at 10, 40 and 70 kOe. In all three images we observe dislocations while the FT shows the existence of orientational order.

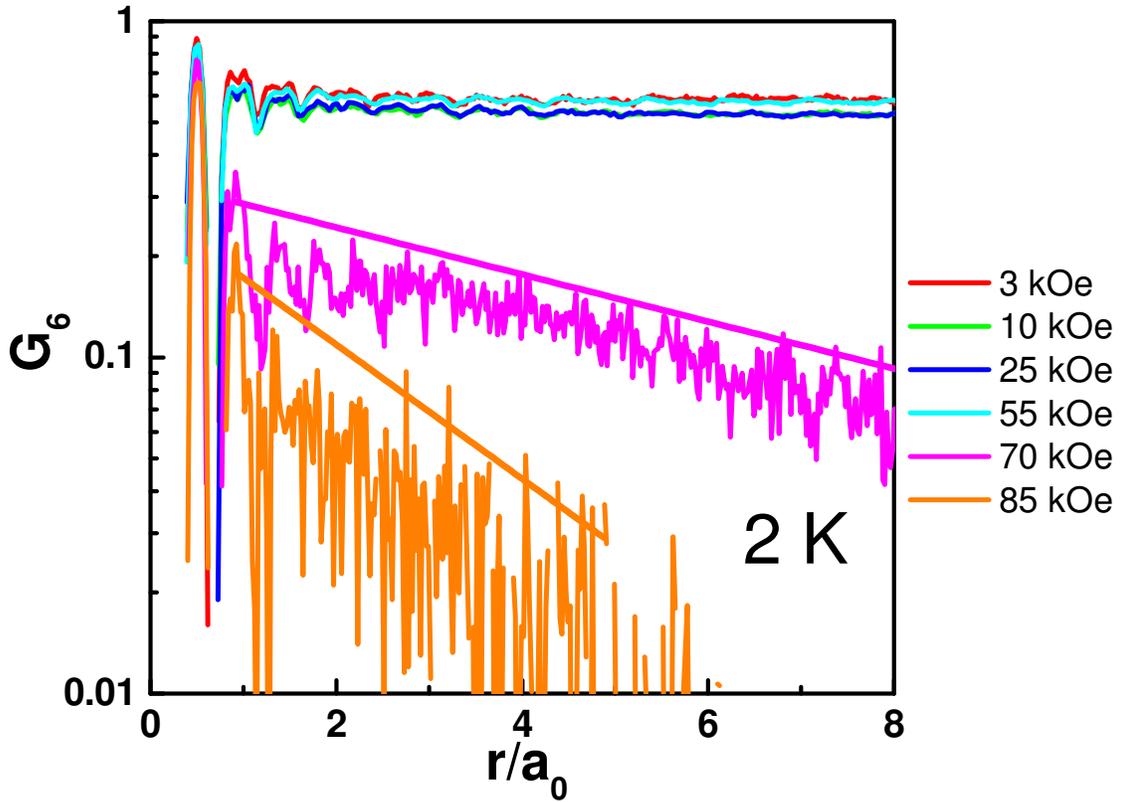

***Figure 4S:*** *$G_6$ as a function of $r/a_0$ for different fields at 2 K. The lines show fits to an exponential decay of the form $G_6 \propto exp(-r/\xi)$ where $\xi$ is the decay length of the orientational order.*

**Section 4: Spatial variation of the orientational order parameter $G_6$**

In the main body of the paper we have discussed the variation of the globally averaged orientational order parameter $\Psi_6$. We can also calculate the spatial decay of orientational order by constructing the metric, $G_6(r) = \langle g_6(0) g_6^*(r) \rangle_{|r|}$, which measures the spatial variation of the local six-fold order parameter $g_6(r) = exp[6i\theta(r)]$, where $\theta(r)$ is the angle of a bond between two nearest-neighbor points on the lattice located at position $r$ with respect to an arbitrary reference axis. For an ideal hexagonal lattice $G_6(r) = 1$. For a hexatic fluid, where the orientational order is quasi-long-range $G_6(r)$ is expected to decrease slowly as a power-law with $r$. In contrast, for a vortex liquid the orientational order is short-range and $G_6(r)$ should decay exponentially. Details of calculating $G_6(r)$ is given is ref. 47.

Figure 4S shows $G_6$ as a function of $r/a_0$, where $a_0$ is the vortex lattice constant, for different fields at 2 K. In the range 10-55 kOe, $G_6 \geq 0.5$ for the largest $r/a_0$ value accessible within the size of our image. Furthermore the slow decay with $r/a_0$ is consistent with the quasi-long-range orientational order in the hexatic phase. (Due to the finite range of $r/a_0$ we cannot determine accurately the power-law exponent of decay in $G_6(r)$.) In contrast, at 70 and 85 kOe the envelope of $G_6$ can be fitted with an exponential decay giving a decay length of $\xi = 6.2a_0$ and $\xi = 2.2a_0$ respectively, as expected for an isotropic vortex liquid.

We can now also understand effect of the finite size of the images on the calculated values of $\Psi_6$. Since at 70 kOe the distance from the center to the edge of our image is of the same order as $\xi$, we obtain a small finite value for $\Psi_6$ (corresponding to the value of $G_6$ at the largest $r/a_0$) even though the orientational order is short-range. Only when $\xi$ becomes much smaller than the image size at 85 kOe we obtain a true zero value.

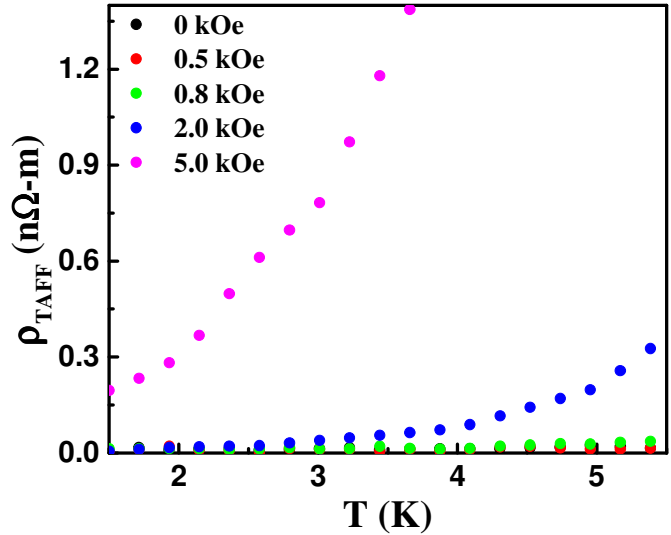

*Figure 5S:* Temperature dependence of $\rho_{TAFF}$ below and above the vortex solid to hexatic fluid transition.

**Section 5: Temperature dependence of $\rho_{TAFF}$ around the solid-hexatic transition**

In Fig.1, we showed that the VL undergoes a transition from a vortex solid to a hexatic fluid at 1.9 kOe. Here we show the temperature dependence of $\rho_{TAFF}$ below and above the transition is also consistent with a vortex solid and a hexatic fluid respectively (Fig. 5S). In the vortex solid state ($H < 1.9$ kOe), as expected, $\rho_{TAFF}$ is zero within the resolution of our measurements over an extended temperature range and becomes finite only at elevated temperatures. In contrast, for H > 1.9 kOe, the resistance increases rapidly from 1.45 K, in qualitative agreement with a vortex fluid state.

**Section 6: Real space imaging of vortex creep**

In Fig. 4 we have shown representative images of vortex creep at 3 different fields. Here we present the complete data at several different fields spanning the the vortex solid, hexatic fluid and isotropic liquid state. In the vortex solid state each vortex undergoes small meandering motion around its mean position and no topological defects appear in the vortex lattice. In the hexatic fluid state the vortices move preferentially along one of the principal direction of the vortex lattice. In the isotropic liquid state the vortex motion is completely random. The creep is visualized by capturing 12 successive images of the vortex lattice at 15 minutes interval. Movies for each of these fields are available as supplementary material.

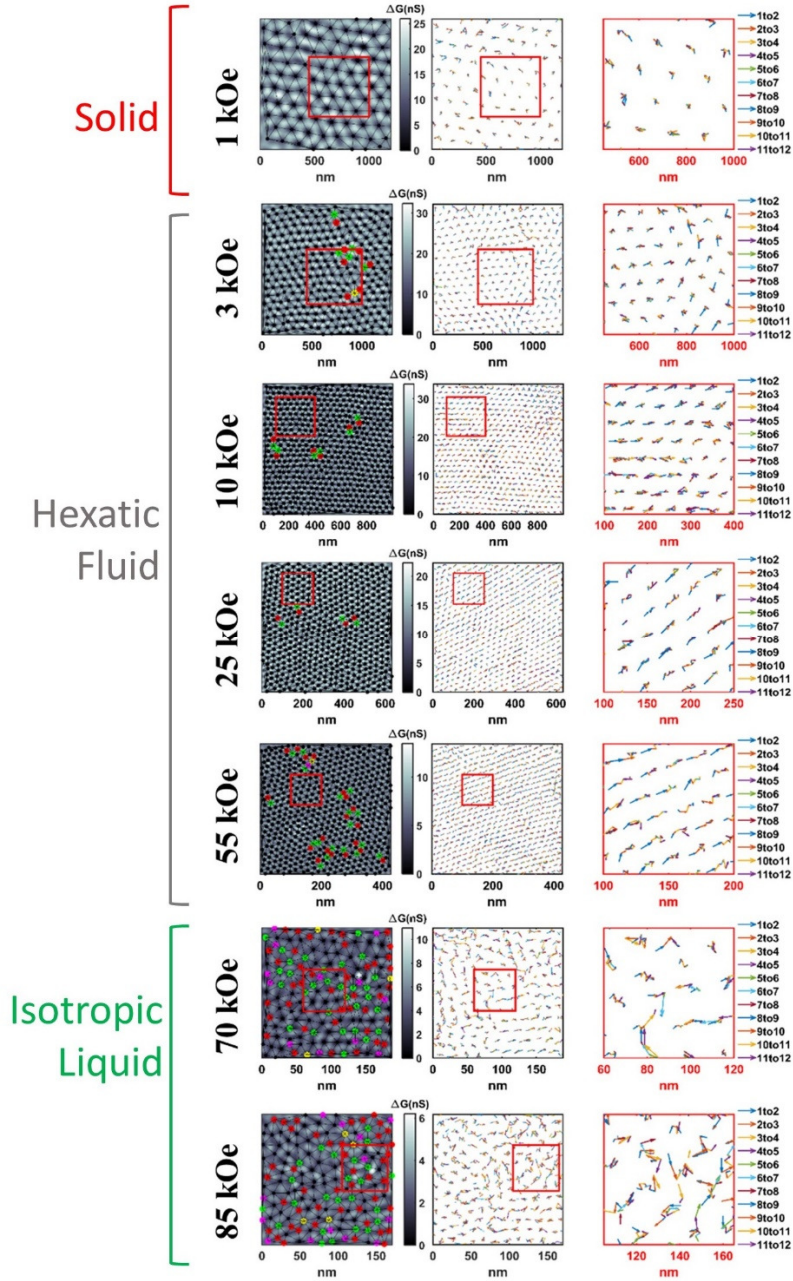

*Figure 6S:* Real space visualization of vortex creep at 2 K in the Vortex Solid, Hexatic Fluid and Isotropic liquid states measured through 12 successive images of the VL taken at 15 minutes interval. The left column shows the first of the 12 successive images of the VL. The middle column shows arrow-maps for each fields, where each arrow gives the displacement for every vortex through individual steps of 12 consecutive vortex images. The right column show the expanded view of the region in the red box in the middle column.

[41] S. J. Turneaure, E. R. Ulm, and T. R. Lemberger, *Numerical modeling of a two-coil apparatus for measuring the magnetic penetration depth in superconducting films and arrays*, J. Appl. Phys. **79**, 4221 (1996).

[42] A. Kamlapure, M. Mondal, M. Chand, A. Mishra, J. Jesudasan, V. Bagwe, L. Benfatto, V. Tripathi and P. Raychaudhuri, *Penetration depth and tunneling studies in very thin epitaxial NbN films,* Appl. Phys. Lett. **96**, 072509 (2010).

[43] I. Roy, P. Chauhan, H. Singh, S. Kumar, J. Jesudasan, P. Parab, R. Sensarma, S. Bose, and P. Raychaudhuri, *Dynamic transition from Mott-like to metal-like state of the vortex lattice in a superconducting film with a periodic array of holes*, Phys. Rev. B **95**, 054513 (2017).

[44] A. Kamlapure, G. Saraswat, S. C. Ganguli, V. Bagwe, P. Raychaudhuri, and S. P. Pai, *A 350 mK, 9 T scanning tunnelling microscope for the study of superconducting thin films on insulating substrates and single crystals*, Rev. Sci. Instrum. **84**, 123905 (2013).

[45] M. Tinkham, *Introduction to Superconductivity, McGraw-Hill Inc. (1996).*

[46] R. C. Dynes, V. Narayanamurti, and J. P. Garno, *Direct Measurement of Quasiparticle-Lifetime Broadening in a Strong-Coupled Superconductor*, Phys. Rev. Lett. **41**, 1509 (1978).

[47] S. C. Ganguli, H. Singh, G. Saraswat, R. Ganguly, V. Bagwe, P. Shirage, A. Thamizhavel and P. Raychaudhuri, *Disordering of the vortex lattice through successive destruction of positional and orientational order in a weakly pinned $Co_{0.0075}NbSe_2$ single crystal*, Sci. Rep. **5**, 10613 (2015).